# Bound states of a quartic and sextic inverse-power-law potential for all angular momenta

A. D. Alhaidari,[(a)] I. A. Assi,[(b)] and A. Mebirouk[(c)]

[(a)] *Saudi Center for Theoretical Physics, P.O. Box 32741, Jeddah 21438, Saudi Arabia*

[(b)] *Department of Physics and Physical Oceanography, Memorial University of Newfoundland, St. John's, Newfoundland & Labrador, A1B 3X7, Canada*

[(c)] *Mathematical Modeling and Numerical Simulation Laboratory, Badji Mokhtar University, BP 12 Annaba, Algeria*

**Abstract:** We use the tridiagonal representation approach to solve the radial Schrödinger equation for an inverse power-law potential of a combined quartic and sextic degrees and for all angular momenta. The amplitude of the quartic singularity is larger than that of the sextic but the signs are negative and positive, respectively. It turns out that the system has a finite number of bound states, which is determined by the larger ratio of the two singularity amplitudes. The solution is written as a finite series of square integrable functions written in terms of the Bessel polynomial.

## 1. Introduction

Power-law potential models have significant applications in various areas of physics, including atomic and molecular physics, particle physics, and gravity physics [1-9]. However, most of these models are exactly solvable if they are less singular than the inverse cube, which is relevant to applications such as electric quadrupole coupling. In this work, we consider the following inverse power-law radial effective potential (including the orbital term) in 3D with spherical symmetry and with quartic and sextic singularities:

$$V(r) = \frac{\ell(\ell+1) + \Lambda}{2r^2} - \frac{b^2}{r^4} + \frac{1}{2}\frac{a^4}{r^6}, \qquad (1)$$

where $\ell$ is the angular momentum quantum number and $\Lambda$ is a dimensionless real parameter. The other length parameters, $a$ and $b$, are real such that $|b| > |a|$. From this point forward, we set $\Lambda = 0$. This potential is highly singular with a finite number of bound states. In the published literature, such potentials are treated only formally. Exact solutions of the Schrödinger equation with this potential for all energies and all angular momenta, which is a highly non-trivial task, has never been reported. In section 2.6 of Ref. [10], Dong studied potential (1) in D dimensions with the extra oscillator term $\frac{1}{2}\omega^2 r^2$ and obtained an energy spectrum proportional to $\omega$ and independent of $\ell$. Therefore, that study gives only a zero energy S-wave solution for potential (1), which is in contradiction to our findings. The Author repeated the same study in 2D with cylindrical symmetry but also found zero energy S-wave solutions [11]. A similar earlier study by Dong and Ma in [12] also resulted in zero energy solutions. Nonetheless, in this work we were able to obtain a solution (energy spectrum and wavefunction) for this potential using the "tridiagonal representation approach" (TRA) (for details of the TRA, one may consult [13] and references therein). The solution of the current problem is written as a finite series of square integrable function written in terms of the Bessel polynomial. The expansion coefficients are orthogonal polynomials in the argument $\left(\ell+\frac{1}{2}\right)^2$ with parameters that depend on the energy and $(b/a)^2$. These polynomials are new and contain all physical information about the system. In sections 2 and 3, we formulate the problem and solve it within the TRA then present the results in sections 4 and 5.

## 2. TRA formulation of the problem

In the atomic units $\hbar = m = 1$, the time-independent radial Schrödinger equation for the inverse power-law potential (1) with $\Lambda = 0$ reads as follows

$$\left[-\frac{1}{2}\frac{d^2}{dr^2} + \frac{\ell(\ell+1)}{2r^2} - \frac{b^2}{r^4} + \frac{1}{2}\frac{a^4}{r^6} - E\right]\psi(r) = 0, \qquad (2)$$

where $E$ is the energy and $\psi(r)$ is the corresponding radial wavefunction. Now, we employ the tools of the TRA in the formulation and solution of this problem (see [13] and references therein). We start by expanding the wavefunction in a series as $\psi(r) = \sum_n f_n \phi_n(x)$, where $\{\phi_n(x)\}$ is a set of square integrable functions that produce a tridiagonal matrix representation for the wave operator (2). We choose the dimensionless variable $x$ as $x = (r/a)^2$. In terms of this variable, the wave equation (2) becomes

$$\mathcal{D}\psi(r) = \frac{-2}{a^2 x}\left[x^2 \frac{d^2}{dx^2} + \frac{x}{2}\frac{d}{dx} - \frac{1}{4}\ell(\ell+1) - \frac{1/4}{x^2} + \frac{(b/a)^2}{2x} + \frac{\varepsilon}{2}x\right]\psi(r) = 0. \qquad (3)$$

where $\varepsilon = a^2 E$. A proper square-integrable basis that could support a tridiagonal matrix representation for the wave operator $\mathcal{D}$ has the following elements

$$\phi_n(x) = A_n x^\alpha e^{-1/2x} Y_n^\mu(x), \qquad (4)$$

where $Y_n^\mu(x)$ is the Bessel polynomial on the positive real line whose relevant properties are given in the Appendix. The normalization constant is conveniently chosen using the orthogonality relation (A3) as $A_n = \sqrt{-(2n+2\mu+1)/n!\Gamma(-n-2\mu)}$. The degree of the polynomial is limited by the negative parameter $\mu$ where $n = 0,1,2,..,N$ with $N$ being the largest integer less than $-\mu - \frac{1}{2}$. Thus, the basis (4) is finite and it cannot faithfully represent continuous scattering states or an infinite number of bound states. However, they are sufficient in describing a finite number of bound states as long as the size of the bound states spectrum is less than or equal to the size of this basis, $N+1$. As we shall see below, the latter is the case for this problem.

Choosing $\alpha = \mu + \frac{3}{4}$ and using the differential equation of the Bessel polynomial (A4), we can evaluate the action of the wave operator on the basis elements giving

$$\mathcal{D}\phi_n(x) = \frac{-A_n}{2a^2} x^{\mu-\frac{1}{4}} e^{-1/2x} \left[ (2n+2\mu+1)^2 - \left(\ell+\tfrac{1}{2}\right)^2 + 2\varepsilon x + \frac{4\mu+2(b/a)^2}{x} \right] Y_n^\mu(x). \quad (5)$$

The TRA dictates that this action takes the following form

$$\mathcal{D}\phi_n(x) = \rho(x)\left[ u_n \phi_n(x) + s_{n-1}\phi_{n-1}(x) + t_n \phi_{n+1}(x) \right], \quad (6)$$

where $\rho(x)$ is an entire function that must be node-less in the open interval $x > 0$. Moreover, it is required that $\{u_n, s_n, t_n\}$ be independent of $x$ with $s_n t_n > 0$ for all $n$. Therefore, a tridiagonal representation that conforms to (6) is obtained if and only if the square brackets in (5) becomes a sum of terms proportional to $Y_n^\mu(x)$ and $Y_{n\pm 1}^\mu(x)$ with constant factors. Hence, the three-term recursion relation (A2) dictates that the terms inside the square brackets is permitted to be only linear in $x$. Thus, the term proportional to $x^{-1}$ must vanish and we should choose the basis parameters $\mu$ as follows

$$2\mu = -(b/a)^2. \quad (7)$$

This is reassuring since the basis parameter $\mu$ is required to be negative. In fact, it should be less than $-\tfrac{1}{2}$, which is consistent with the physical requirement that $|b| > |a|$. Moreover, since $N$ is the largest integer less than $-\mu - \tfrac{1}{2}$, then the maximum number of bound states that could be obtained by our TRA solution is

$$N_{max} + 1 = \left\lfloor \tfrac{1}{2}(b/a)^2 + \tfrac{1}{2} \right\rfloor, \quad (8)$$

which is independent of the angular momentum. However, the actual size of the bound states spectrum does depend on the angular momentum as will be confirmed by our results below. In fact, physically we expect that the number of bound states to decrease with increasing $\ell$ where the disappearing bound states turn into resonances [14]. Therefore, the maximum number of bound states given by Eq. (8) corresponds to $\ell = 0$. In the following section, we present the TRA solution of the problem which is written in terms of a new polynomial defined in [15] by its three-term recursion relation and initial values.

## 3. TRA solution of the problem

With the basis parameters $\mu$ given by (7), the action of the wave operator on the basis elements (4) reduces to the following

$$\mathcal{D}\phi_n(x) = \frac{-A_n}{2a^2} x^{\mu-\frac{1}{4}} e^{-1/2x} \left[ (2n+2\mu+1)^2 - \left(\ell+\tfrac{1}{2}\right)^2 + 2\varepsilon x \right] Y_n^\mu(x). \quad (9)$$

Substituting this action in the wave equation $\mathcal{D}\psi(r) = \sum_n f_n \mathcal{D}\phi_n(x) = 0$ and using the recursion relation of the Bessel polynomial (A2), we obtain the following three-term recursion relation for the expansion coefficients

$$\left(\ell+\tfrac{1}{2}\right)^2 P_n = \left[(2n+2\mu+1)^2 - \frac{\mu\varepsilon}{(n+\mu)(n+\mu+1)}\right]P_n$$
$$+\varepsilon\left[\frac{1}{n+\mu+1}\sqrt{\frac{-(n+1)(n+2\mu+1)}{(2n+2\mu+1)(2n+2\mu+3)}}P_{n+1} + \frac{1}{n+\mu}\sqrt{\frac{-n(n+2\mu)}{(2n+2\mu-1)(2n+2\mu+1)}}P_{n-1}\right] \quad (10)$$

where we have written $f_n = f_0 P_n$ making $P_0 = 1$. This recursion relation gives $P_n$ as a polynomial of degree $n$ in $\left(\ell+\tfrac{1}{2}\right)^2$ with parameters that depend on the energy $\varepsilon$ and $\mu$. Comparing this recursion relation to that of the polynomial $B_n^\mu(z;\gamma)$ defined in [15] by its three-term recursion relation, which is shown here in the Appendix as (A10), we conclude that $P_n = A_n B_n^\mu(z;\gamma)$ with

$$\gamma = 8/\varepsilon, \qquad z = (2/\varepsilon)\left(\ell+\tfrac{1}{2}\right)^2. \qquad (11)$$

Finally, the kth bound state wavefunction for this power-law potential reads

$$\psi_k(r) = f_0(E_k)\left(r^2/a^2\right)^{\mu+\tfrac{3}{4}} e^{-a^2/2r^2} \sum_{n=0}^{N} A_n^2 B_n^\mu(z;\gamma) Y_n^\mu(r^2/a^2). \qquad (12)$$

For a given set of physical parameters $\{a,b,\ell\}$ and bound state energy $E_k$, the basis parameter $\mu$ is given by (7) whereas $z$ and $\gamma$ are given by (11). All properties of the system (e.g., energy spectrum of the bound states, density of states, etc.) are obtained from the properties of the polynomial $B_n^\mu(z;\gamma)$ (e.g., its weight function, generating function, asymptotics, zeros, etc.). For example, the energy spectrum could easily be obtained from the spectrum formula of this polynomial. Unfortunately, the analytic properties of $B_n^\mu(z;\gamma)$ are not yet known. It remains an open problem in orthogonal polynomials along with other similar problems (for an expose of these problems, one may consult [16] and references therein). Therefore, we are forced to resort to numerical means to calculate the physical properties of the system. The most important property needed for our system is the energy spectrum $\{E_k\}$ because then the corresponding wavefunction (12) will be fully determined. This will be done in the following section where we obtain the bound states energy spectrum for a given set of physical parameters $\{a,b,\ell\}$.

## 4. The energy spectrum

To find the energy spectrum numerically in the absence of the analytic properties of the TRA polynomial $B_n^\mu(z;\gamma)$, we might be tempted to use the recursion relation (10) written as an eigenvalue equation $q|P\rangle = T|P\rangle$ where $q = \left(\ell+\tfrac{1}{2}\right)^2$ and $T$ is the symmetric tridiagonal matrix $T_{n,m} = A_n \delta_{n,m} + B_{n-1} \delta_{n,m+1} + B_n \delta_{n,m-1}$ with

$$A_n = (2n+2\mu+1)^2 + \frac{-\varepsilon\mu}{(n+\mu)(n+\mu+1)}, \qquad (13a)$$

$$B_n = \frac{\varepsilon}{n+\mu+1}\sqrt{\frac{-(n+1)(n+2\mu+1)}{(2n+2\mu+1)(2n+2\mu+3)}}. \qquad (13b)$$

However, the problem in finding the energy spectrum this way is that the matrix elements of $T$ do depend on the energy itself that we are trying to find. Of course, out of all energies that enter in the construction of $T$, the discrete energies that belong to the spectrum are only those that produce the eigenvalue $q = \left(\ell + \tfrac{1}{2}\right)^2$ for a given angular momentum $\ell$. Therefore, we choose an alternative numerical technique to accomplish this task. It goes as follows: The action of the Hamiltonian operator on the basis elements, $H|\phi_n\rangle$, is obtained from (9) by setting $\varepsilon = 0$. Subsequently, the Hamiltonian matrix elements $\langle \phi_m | H | \phi_n \rangle$ are obtained from that using the orthogonality relation (A3) together with the integration measure $a^{-1}\int_0^\infty (...)dr = 2^{-1}\int_0^\infty x^{-1/2}(...)dx$. The result is an $(N+1)\times(N+1)$ diagonal matrix with the following elements

$$\langle \phi_n | H | \phi_m \rangle = \frac{1}{4a^2}\left[\left(\ell + \tfrac{1}{2}\right)^2 - (2n + 2\mu + 1)^2\right]\delta_{n,m}, \tag{14}$$

where $n, m = 0, 1, 2, .., N$. These diagonal elements would have been identified as the energy eigenvalues had the basis set (4) been orthogonal; but it is not. The energy spectrum is obtained by solving the generalized eigenvalue equation $H|\psi\rangle = E\Omega|\psi\rangle$, where $\Omega$ is the basis overlap matrix whose elements are $\Omega_{n,m} = \langle \phi_n | \phi_m \rangle \neq \delta_{n,m}$. Using the orthogonality (A3) and the recursion relation (A2) together with the integration measure noted above, we obtain $\Omega$ as the following $(N+1)\times(N+1)$ tridiagonal symmetric matrix

$$\Omega_{n,m} = \frac{1}{4}\left(\tilde{A}_n \delta_{n,m} + \tilde{B}_{n-1}\delta_{n,m+1} + \tilde{B}_n \delta_{n,m-1}\right). \tag{15}$$

where $\tilde{A}_n$ and $\tilde{B}_n$ are the factors that multiply $\varepsilon$ in $A_n$ and $B_n$ of Eq. (13), respectively. Now, with the matrices $H$ and $\Omega$ being determined, we can obtain the energy spectrum of the bound states as the negative subset of the $N+1$ eigenvalues of the generalized eigenvalue equation $H|\psi\rangle = E\Omega|\psi\rangle$. These are shown in Table 1 for a given set of physical parameters $\{a,b\}$ and several values of the angular momentum. Since this energy spectrum is obtained numerically using finite matrices whose size, which could be very small, is dictated by the physical parameters $a$ and $b$, then it would be desirable to obtain an independent verification of these results using other numerical means. This is carried out in the following section where we obtain the energy spectrum by two different numerical methods.

## 5. Results and discussion

In the first method, we select an energy independent complete square integrable basis in which the matrix representation of the Hamiltonian is tridiagonal and symmetric. Then, we diagonalize a finite submatrix of this Hamiltonian for a given set of physical parameters $\{a,b,\ell\}$ and obtain the corresponding energy spectrum as the negative eigenvalues of this finite submatrix. We will also demonstrate that the spectrum converges rapidly with an increase in the size of the submatrix. Moreover, we compare these results with those obtained by higher order finite difference schemes for solving the eigenvalue problem [19].

For the complete square integrable basis, we choose the "Laguerre basis" whose elements read as follows

$$\chi_n(y) = C_n y^\alpha e^{-y/2} L_n^\nu(y), \tag{16}$$

where $y = (\lambda r)^{-2}$ and $L_n^\nu(y)$ is the Laguerre polynomial with $\lambda$ being a positive scale parameter of inverse length dimension and $\nu > -1$. The normalization constant is chosen as $C_n = \sqrt{2(n!)/\Gamma(n+\nu+1)}$. Taking $\nu = \ell + \frac{1}{2}$ and $\alpha = \frac{1}{2}\ell$, the action of the Hamiltonian operator, $H = -\frac{1}{2}\frac{d^2}{dr^2} + V(r)$, on the basis (16) becomes

$$H\chi_n(y) = \lambda^2 C_n y^{\alpha+2} e^{-y/2}\left\{(2n+\nu+1) - (\lambda b)^2 + \frac{1}{2}\left[(\lambda a)^4 - 1\right]y\right\} L_n^\nu(y), \tag{17}$$

where we have used $\frac{d^2}{dr^2} = 4\lambda^2 y^2\left(y\frac{d^2}{dy^2} + \frac{3}{2}\frac{d}{dy}\right)$ and utilized the differential equation of the Laguerre polynomial,

$$\left[y\frac{d^2}{dy^2} + (\nu+1-y)\frac{d}{dy} + n\right]L_n^\nu(y) = 0. \tag{18}$$

Now, we use the recursion relation of the Laguerre polynomials, $yL_n^\nu(y) = (2n+\nu+1)L_n^\nu(y) - (n+\nu)L_{n-1}^\nu(y) - (n+1)L_{n+1}^\nu(y)$ for the last term inside the curly brackets in (17). The result is as follows

$$H\chi_n(y) = \frac{\lambda^2}{2} y^2\left\{\left[(\lambda a)^4 + 1\right](2n+\nu+1) - 2(\lambda b)^2\right\}\chi_n(y)$$
$$-\frac{\lambda^2}{2} y^2\left[(\lambda a)^4 - 1\right]\left\{\sqrt{n(n+\nu)}\,\chi_{n-1}(y) + \sqrt{(n+1)(n+\nu+1)}\,\chi_{n+1}(y)\right\} \tag{19}$$

Using the integration measure $\lambda\int_0^\infty (...)dr = -\frac{1}{2}\int_\infty^0 y^{-3/2}(...)dy$ and the orthogonality relation of the Laguerre polynomials, $\int_0^\infty y^\nu e^{-y} L_n^\nu(y) L_m^\nu(y) dx = \frac{\Gamma(n+\nu+1)}{\Gamma(n+1)}\delta_{n,m}$, we obtain the following tridiagonal symmetric Hamiltonian matrix

$$\langle\chi_n|H|\chi_m\rangle = \frac{\lambda^2}{2}\left\{\left[(\lambda a)^4 + 1\right]\left(2n+\ell+\tfrac{3}{2}\right) - 2(\lambda b)^2\right\}\delta_{n,m}$$
$$-\frac{\lambda^2}{2}\left[(\lambda a)^4 - 1\right]\left\{\sqrt{n\left(n+\ell+\tfrac{1}{2}\right)}\,\delta_{n,m+1} + \sqrt{(n+1)\left(n+\ell+\tfrac{3}{2}\right)}\,\delta_{n,m-1}\right\} \tag{20}$$

Therefore, the energy spectrum $\{E_k\}_{k=0}^K$ is obtained as the $K+1$ negative eigenvalues of the following generalized eigenvalue equation

$$\mathcal{H}|\psi_k\rangle = E_k \Omega|\psi_k\rangle, \tag{21}$$

where $\mathcal{H}$ is the Hamiltonian matrix whose elements are given by (20) and $\Omega$ is the overlap matrix of the basis (16) whose elements are

$$\Omega_{n,m} = \langle \chi_n | \chi_m \rangle = \lambda \int_0^\infty \chi_n(y) \chi_m(y) dr = \frac{C_n C_m}{2} \int_0^\infty y^{\nu-2} e^{-y} L_n^\nu(y) L_m^\nu(y) dy. \qquad (22)$$

This integral could be evaluate numerically using Gauss quadrature integral approximation associated with the Laguerre polynomials (see, for example, Appendix B in [20]). Rigorously, the integrability of (22) dictates that the exponent $\nu - 2$ of $y$ must be greater than $-1$. Thus, with $\nu = \ell + \frac{1}{2}$, we expect integration difficulty only for $\ell = 0$. However, numerically we can improve the accuracy of the result of integration by increasing the size of the basis (16) and choosing a proper value of the scale parameter $\lambda$. For a given choice of the physical parameters $\{a,b\}$ and several angular momenta, Table 2 gives the complete bound states energies. One should note that the overlap matrix $\Omega$ does not depend on the non-physical computational scale parameter $\lambda$ but the Hamiltonian matrix $\mathcal{H}$ does. However, this dependence is a numerical artifact in the sense that the computed energy eigenvalues $\{E_k\}_{k=0}^K$ will be independent of $\lambda$ if the size of the matrices becomes very large. For finite size matrices, though, we search for a range of values of $\lambda$ where the eigenvalues do not change (within the desired accuracy) as we vary $\lambda$ within this range. This range is called the "plateau of stability" for $\lambda$ whose width increases with the size of the matrices. Theoretically, the width of the plateau goes to infinity as the size of the matrices go to infinity. Our choice of $\lambda$ for all results in this work is picked up from the middle of the plateau. Table 3 shows good convergence of the energy spectrum as the size of the matrices $\mathcal{H}$ and $\Omega$ increases.

For the second method, which is based on the finite difference scheme [19], we start by applying a change of variable $s = \frac{2}{\pi} \tan^{-1}(\tau r)$ to the Schrödinger equation (2), giving

$$\frac{2\tau^2}{\pi^2} \left\{ -\cos^4\left(\frac{\pi}{2}s\right) \frac{d^2}{ds^2} + \pi \cos^3\left(\frac{\pi}{2}s\right) \sin\left(\frac{\pi}{2}s\right) \frac{d}{ds} + \frac{\pi^2}{2\tau^2}[V(s) - E] \right\} \psi(s) = 0, \qquad (23)$$

where $\tau$ is some real positive parameter of inverse length dimension to be fixed later and

$$V(s) = \frac{\tau^2/2}{\tan^2\left(\frac{\pi}{2}s\right)} \left[ \ell(\ell+1) - \frac{2(b\tau)^2}{\tan^2\left(\frac{\pi}{2}s\right)} + \frac{(a\tau)^4}{\tan^4\left(\frac{\pi}{2}s\right)} \right], \qquad (24)$$

Next, we discretize the finite interval $[0,1]$ into a uniform grid, $s_i = ih$, where $i = 0, 1, ..., M+1$ and $h = 1/(M+1)$ for some large enough integer $M$. Next, we apply $2k$-step finite difference scheme to approximate the derivatives in Eq. (23) [19]

$$\left. \frac{d\psi}{ds} \right|_{s_i} \approx \frac{1}{h} \begin{cases} \sum_{j=0}^{2k} \delta_{i,j,1} \psi_j, & i = 1, 2, 3, ..., k-1 \\ \sum_{j=0}^{2k} \delta_{k,j,1} \psi_{i+j-k}, & i = k, k+1, ..., M+1-k \\ \sum_{j=0}^{2k} \delta_{i-t,j,1} \psi_{j+t}, & i = M+2-k, ..., M \ \& \ t = M-2k+1 \end{cases} \qquad (25)$$

$$\left.\frac{d^2\psi}{ds^2}\right|_{s_i} \approx \frac{1}{h^2}\begin{cases} \sum_{j=0}^{2k+1}\delta_{i,j,2}\psi_j, & i=1,2,3,\ldots,k-1 \\ \sum_{j=0}^{2k}\delta_{k,j,2}\psi_{i+j-k}, & i=k,k+1,\ldots,M+1-k \\ \sum_{j=0}^{2k+1}\delta_{k,j,2}\psi_{j+t-1}, & i=M+2-k,\ldots,M \ \& \ t=M-2k+1 \end{cases} \quad (26)$$

where $\{\delta_{i,j,l}\}$ with $l=1,2$ are the weights which are determined by requiring the formula to have maximum order of consistency (see [19] for more details), and $\psi_j = \psi(s_j)$. Using the separated homogeneous Dirichlet boundary condition $\psi_0 = \psi(s_0) = 0$ and $\psi_{M+1} = \psi(s_{M+1}) = 0$ along with the above approximation schemes, we transform Eq. (3) to the following eigenvalue problem,

$$J\Psi = (A\Delta_2 + B\Delta_1 + CI_M)\Psi = E\Psi, \quad (27)$$

where $A$, $B$ and $C$ are diagonal matrices of size $M$ which contain the coefficients in Eq. (23). $\Delta_l$ with $l=1,2$ are real square matrices of size $M$ which contain the coefficients $\delta_{i,j,l}/h^l$. $I_M$ is the $M \times M$ identity matrix and $\Psi = (\psi_1, \psi_2, \ldots, \psi_M)^T$. The parameter $\tau$ is chosen as $\tau = 0.6(j)^{-0.7}$, where $j = 1, \ldots, M$ is the eigenvalue index [19]. Therefore, the complete solutions of the Schrödinger equation (2) are approximated by the eigenvalues and the corresponding eigenvectors of the matrix $J$. In Table 4, we have computed the eigenvalues of $J$ for the given choice of parameters using "Matlab" which is in good agreement with the results obtained by the Hamiltonian diagonalization method in the Laguerre basis (Table 2). The discrepancy between the results in Tables 1 and those in Table 2 or 4 is expected due to the finite size basis used in producing Table 1 as explained at the end of Sec. 4.

Figure 1 is a plot of the energy spectrum for the lowest states as a function of the angular momentum. In an attempt to discover the energy spectrum formula, we used the **linfit** built-in function in Mathcad that fits data points to a properly chosen function suggested by the shapes of the curves in Figure 1. Accordingly, we obtained the following possible energy spectrum formula

$$E(n,\ell) = C_2(n)\ell^2 - C_0(n), \quad (28)$$

where the two functions $C_0(n)$ and $C_2(n)$ are positive that also depend on the physical parameters $\{a,b\}$. Moreover, the figure suggests that $\frac{d}{dn}C_2(n)$ is negative. Figure 2 is a plot of the energy spectrum $\{E_n\}$ as a function of $n$ for several angular momenta. Figure 3 is a plot of the bound states wavefunctions corresponding to the $\ell = 3$ column in Table 2.

## 6. Conclusion

In this work, we obtained a finite exact series solution of the Schrödinger equation for the bound states of a radial inverse power-law potential with quartic and sextic singularities. The solution series is written in terms of square integrable functions using the Bessel polynomial. The expansion coefficients are orthogonal polynomials in the energy, angular momentum and potential parameters. We used the tridiagonal representation approach as the method of solution and verified our results numerically via two independent computational methods. We expect that the solution obtained here will have fruitful applications in areas where inverse power-law potential models are relevant such as in molecular physics and high energy physics.

## Appendix: Bessel polynomial on the real line

The Bessel polynomial on the positive real line is defined in terms of the hypergeometric or confluent hypergeometric functions as follows (see section 9.13 of the book by Koekoek *et. al* [21] but make the replacement $x \mapsto 2x$ and $a \mapsto 2\mu$)

$$Y_n^\mu(x) = {}_2F_0\left(\begin{matrix}-n, n+2\mu+1\\ -\end{matrix}\middle|-x\right) = (n+2\mu+1)_n x^n {}_1F_1\left(\begin{matrix}-n\\ -2(n+\mu)\end{matrix}\middle|1/x\right),\tag{A1}$$

where $x \geq 0$, $n = 0,1,2,...,N$ and $N$ is a non-negative integer. The real parameter $\mu$ is negative such that $\mu < -N - \frac{1}{2}$. The Pochhammer symbol $(a)_n$ (a.k.a. shifted factorial) is defined as $(a)_n = a(a+1)(a+2)...(a+n-1) = \frac{\Gamma(n+a)}{\Gamma(a)}$. The Bessel polynomial could also be written in terms of the associated Laguerre polynomial as: $Y_n^\mu(x) = n!(-x)^n L_n^{-(2n+2\mu+1)}(1/x)$. The three-term recursion relation reads as follows:

$$2x Y_n^\mu(x) = \frac{-\mu}{(n+\mu)(n+\mu+1)} Y_n^\mu(x)$$
$$-\frac{n}{(n+\mu)(2n+2\mu+1)} Y_{n-1}^\mu(x) + \frac{n+2\mu+1}{(n+\mu+1)(2n+2\mu+1)} Y_{n+1}^\mu(x)\tag{A2}$$

Note that the constraints on $\mu$ and on the maximum polynomial degree make this recursion definite (i.e., the signs of the two recursion coefficients multiplying $Y_{n\pm1}^\mu(x)$ are the same). Otherwise, these polynomials could not be defined on the real line but on the unit circle in the complex plane. The orthogonality relation reads as follows

$$\int_0^\infty x^{2\mu} e^{-1/x} Y_n^\mu(x) Y_m^\mu(x) dx = -\frac{n!\Gamma(-n-2\mu)}{2n+2\mu+1}\delta_{nm}.\tag{A3}$$

The differential equation is

$$\left\{x^2\frac{d^2}{dx^2} + [1+2x(\mu+1)]\frac{d}{dx} - n(n+2\mu+1)\right\}Y_n^\mu(x) = 0.\tag{A4}$$

The forward and backward shift differential relations read as follows

$$\frac{d}{dx}Y_n^\mu(x) = n(n+2\mu+1)Y_{n-1}^{\mu+1}(x).\tag{A5}$$

$$x^2\frac{d}{dx}Y_n^\mu(x) = -(2\mu x+1)Y_n^\mu(x) + Y_{n+1}^{\mu-1}(x).\tag{A6}$$

We can write $Y_{n+1}^{\mu-1}(x)$ in terms of $Y_n^\mu(x)$ and $Y_{n\pm1}^\mu(x)$ as follows

–9–

$$2Y_{n+1}^{\mu-1}(x) = \frac{(n+1)(n+2\mu)}{(n+\mu)(n+\mu+1)} Y_n^{\mu}(x)$$
$$+ \frac{n(n+1)}{(n+\mu)(2n+2\mu+1)} Y_{n-1}^{\mu}(x) + \frac{(n+2\mu)(n+2\mu+1)}{(n+\mu+1)(2n+2\mu+1)} Y_{n+1}^{\mu}(x) \quad (A7)$$

Using this identity and the recursion relation (A2), we can rewrite the backward shift differential relation as follows

$$2x^2 \frac{d}{dx} Y_n^{\mu}(x) = n(n+2\mu+1) \times$$
$$\left[ -\frac{Y_n^{\mu}(x)}{(n+\mu)(n+\mu+1)} + \frac{Y_{n-1}^{\mu}(x)}{(n+\mu)(2n+2\mu+1)} + \frac{Y_{n+1}^{\mu}(x)}{(n+\mu+1)(2n+2\mu+1)} \right] \quad (A8)$$

The generating function is

$$\sum_{n=0}^{\infty} Y_n^{\mu}(x) \frac{t^n}{n!} = \frac{2^{2\mu}}{\sqrt{1-4xt}} \left(1+\sqrt{1-4xt}\right)^{-2\mu} \exp\left[2t/(1+\sqrt{1-4xt})\right]. \quad (A9)$$

The polynomial $B_n^{\mu}(z;\gamma)$ is defined in [15] by its three-term recursion relation Eq. (16) therein, which reads

$$z B_n^{\mu}(z;\gamma) = \left[ \frac{-2\mu}{(n+\mu)(n+\mu+1)} + \gamma\left(n+\mu+\tfrac{1}{2}\right)^2 \right] B_n^{\mu}(z;\gamma)$$
$$- \frac{n}{(n+\mu)\left(n+\mu+\tfrac{1}{2}\right)} B_{n-1}^{\mu}(z;\gamma) + \frac{n+2\mu+1}{(n+\mu+1)\left(n+\mu+\tfrac{1}{2}\right)} B_{n+1}^{\mu}(z;\gamma) \quad (A10)$$

where $B_0^{\mu}(z;\gamma) = 1$ and $B_{-1}^{\mu}(z;\gamma) := 0$.

## Tables Caption:

**Table 1**: The complete bound states energy spectrum (in atomic units with an overall negative sign) for several angular momenta. We took $a = 2$ and $b = 7$.

**Table 2**: Reproduction of Table 1 using numerical diagonalization of the Hamiltonian matrix in the basis (16). The Hamiltonian matrix size is $100 \times 100$ and the values of scale parameter $\lambda$ are as shown.

**Table 3**: Reproduction of the first column of Table 2 for different matrix sizes. Good convergence with increasing matrix size is evident. The physical parameters are $a = 2$, $b = 7$ and $\ell = 0$.

**Table 4**: Reproduction of Table 1 using the 16$^{th}$ order ($k = 8$) finite difference with grid size $M = 1000$. [19].

## Figures Caption:

**Fig. 1**: The bound states energy spectrum (in atomic units) as a function of the angular momentum. The energy levels $n = (0,1,2,..,10)$ correspond to the curves from bottom to top. The values of the physical parameters are $a = 2$ and $b = 15$.

**Fig. 2**: The bound states energy spectrum (in atomic units) as a function of the level quantum number $n$. The angular momenta $\ell = (0,10,20,30,40,50)$ correspond to the curves from bottom to top. The values of the physical parameters are $a = 2$ and $b = 15$.

**Fig. 3**: Plots of the un-normalized bound states wavefunctions corresponding to the $\ell = 3$ column in Table 2.

**Table 1**

| n | $\ell=0$ | $\ell=1$ | $\ell=2$ | $\ell=3$ | $\ell=4$ | $\ell=5$ |
|---|---|---|---|---|---|---|
| 0 | 195.833847586 | 192.405028790 | 185.564643590 | 175.349754004 | 161.822874281 | 145.081815855 |
| 1 | 90.848444079 | 88.568922492 | 84.011634455 | 77.183141968 | 68.103260502 | 56.825420015 |
| 2 | 33.647121748 | 32.322076951 | 29.665398497 | 25.665506229 | 20.314596934 | 13.650143026 |
| 3 | 8.633176082 | 8.001943877 | 6.731617802 | 4.806042927 | 2.219811322 | |
| 4 | 1.041599359 | 0.842693957 | 0.440432619 | | | |
| 5 | 0.007529895 | | | | | |

**Table 2**

| n | $\ell=0$ ($\lambda=0.25$) | $\ell=1$ ($\lambda=0.30$) | $\ell=2$ ($\lambda=0.25$) | $\ell=3$ ($\lambda=0.30$) | $\ell=4$ ($\lambda=0.25$) | $\ell=5$ ($\lambda=0.35$) |
|---|---|---|---|---|---|---|
| 0 | 198.053812 | 194.485062 | 187.382284 | 176.816075 | 162.895376 | 145.771664 |
| 1 | 95.107392 | 92.507287 | 87.351047 | 79.728921 | 69.782625 | 57.714758 |
| 2 | 37.503431 | 35.790301 | 32.419721 | 27.507836 | 21.242776 | 13.908937 |
| 3 | 10.649435 | 9.698773 | 7.867300 | 5.305988 | 2.281243 | |
| 4 | 1.563860 | 1.193393 | 0.540802 | | | |
| 5 | 0.025307 | | | | | |

**Table 3**

| n | 30×30 | 40×40 | 50×50 | 70×70 | 100×100 |
|---|---|---|---|---|---|
| 0 | 197.492537252 | 198.052373573 | 198.053810517 | 198.053811738 | 198.053811718 |
| 1 | 94.314048133 | 95.104944526 | 95.107389328 | 95.107391512 | 95.107391514 |
| 2 | 36.915040547 | 37.501405664 | 37.503429021 | 37.503430893 | 37.503430893 |
| 3 | 10.375623268 | 10.648439288 | 10.649434217 | 10.649435153 | 10.649435153 |
| 4 | 1.494065003 | 1.563599312 | 1.563859420 | 1.563859664 | 1.563859663 |
| 5 | 0.021836656 | 0.025369900 | 0.025380133 | 0.025339205 | 0.025307315 |

**Table 4**

| n | $\ell = 0$ | $\ell = 1$ | $\ell = 2$ | $\ell = 3$ | $\ell = 4$ | $\ell = 5$ |
|---|---|---|---|---|---|---|
| 0 | 198.053811725 | 194.485061786 | 187.382284129 | 176.816074818 | 162.895376078 | 145.771664441 |
| 1 | 95.107391509 | 92.507287345 | 87.351046558 | 79.728921145 | 69.782624776 | 57.714757854 |
| 2 | 37.503430893 | 35.790300903 | 32.419721196 | 27.507836187 | 21.242775651 | 13.908937252 |
| 3 | 10.649435153 | 9.698773354 | 7.867299662 | 5.305987630 | 2.281242651 | |
| 4 | 1.563859663 | 1.193393381 | 0.540802068 | | | |
| 5 | 0.025300267 | | | | | |

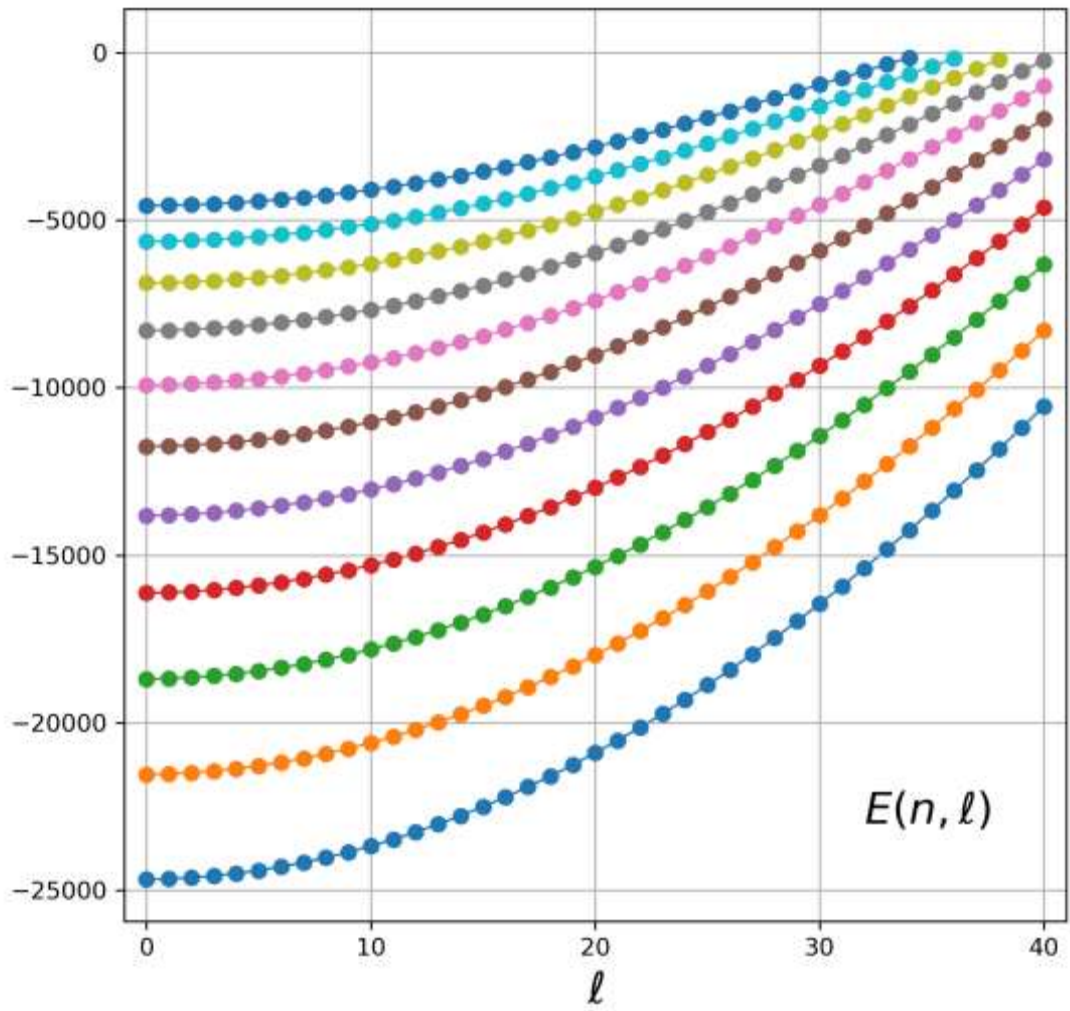

**Fig. 1**

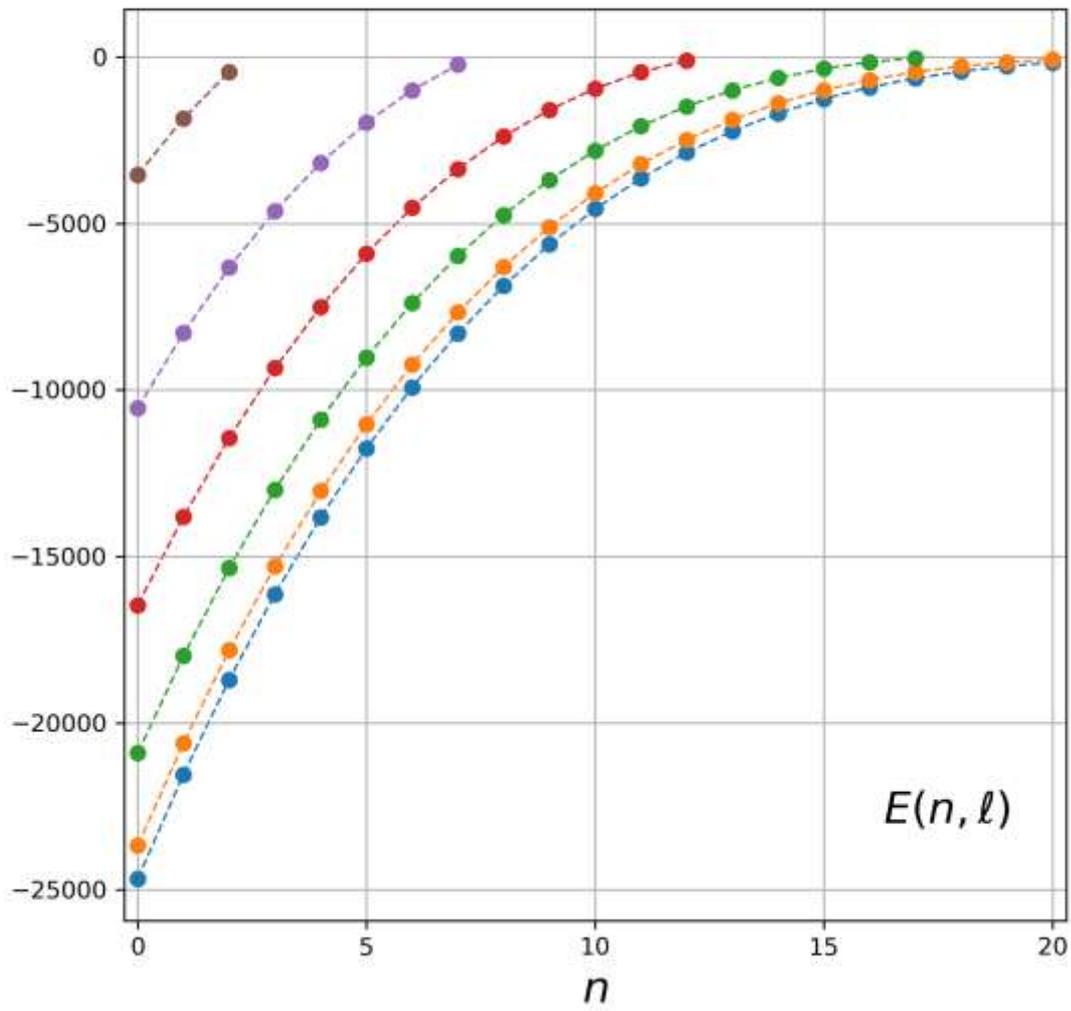

**Fig. 2**

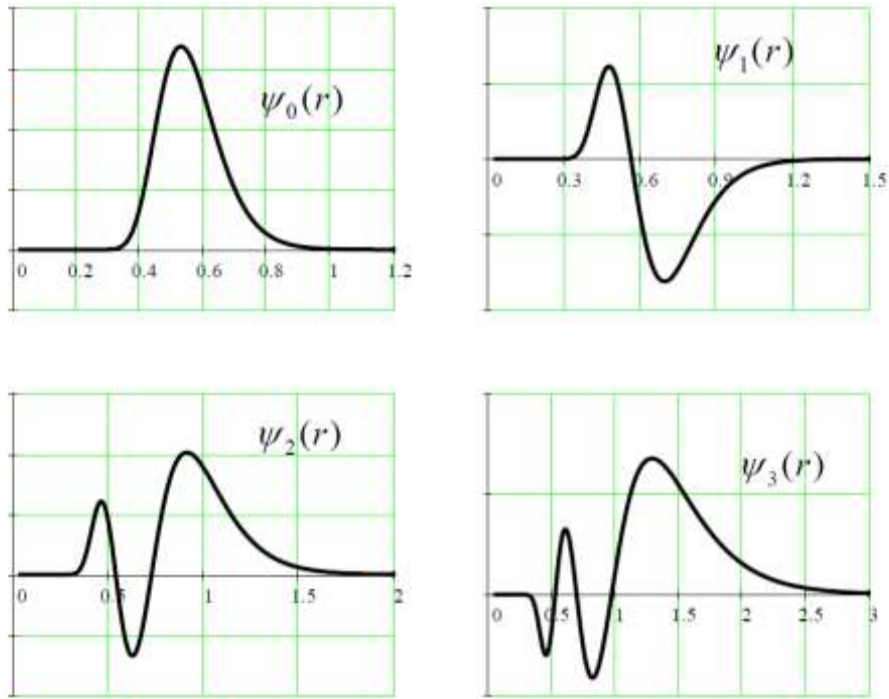

**Fig. 3**